# Determining the most efficient geometry through simulation study of ZnO nanorods for the development of high performance tactile sensors and energy harvesting devices

*Rehan Ahmed and Pramod Kumar\**

Department of Physics, Indian Institute of Technology Bombay, Mumbai, 400076, India

*Corresponding author, *E-mail: pramod_k@iitb.ac.in

## Abstract

The piezoelectric nanomaterial ZnO exhibits an excellent piezoelectric response that can transduce mechanical energy into electrical signals by applying pressure. The piezoelectric behavior of ZnO nanostructures (especially nanorods or microrods) is getting considerable attention in the fabrications of piezo tactile sensors, energy harvesting devices, and other self-powering implantable devices. Especially vertically aligned ZnO nanorods are of high interest due to their higher value of piezoelectric coefficient along the z-direction. In this report, various geometries and alignments of ZnO nanorods are explored and their effect on strength of piezoelectric output potential has been simulated by COMSOL Multiphysics software. Best suited geometry and inclination are explored in this simulation to achieve high piezoelectric output in haptic and energy harvester devices. The simulation results show out of many geometries and inclinations the highest piezoelectric output is demonstrated by the inclined ZnO nanorods due to the application of higher torque force or shear stress in similar applied force. The high torque force or shear stress at 60° orientation and optimized contributions from all the piezoelectric coefficients resulted in high piezoelectric output potential close to 215 mV which is much higher than the vertically aligned

ZnO nanorod which is approximately 25 mV. The results are contrary to accepted understanding that the vertical ZnO nanorods should produce highest output voltage due to high piezoelectric coefficient along z-axis.

**Keywords;** ZnO nanorods, simulation, piezoelectricity

## 1. Introduction

Micro-electromechanical self-powered devices such as piezo tactile sensors,[1] electronic skin,[2] nano field-effect transistor (FET),[3] functional nanogenerator,[4] and energy harvesting nano devices,[5] are inexorably dependent on emerging piezoelectric materials. These devices primarily require piezoelectric material which can be in various forms from thin films to monocrystalline forms. The rapid progress of piezoelectric materials in nanotechnology has received a great deal of attention due to its various applications. Piezoelectricity is an electromechanical process in which an electric dipole is generated due to the deformation of the piezoelectric material by applying a mechanical force or pressure, called the direct piezoelectric effect. While the inverse piezoelectric effect refers to applying an electric field in the polarization direction of piezoelectric materials, and it will produce a mechanical deformation in the form of strain. There are various piezoelectric materials such as quartz,[6] aluminum nitride,[7] lead zirconate titanate (PZT),[8] and wurtzite zinc oxide (ZnO),[9,10,11] that play an important and key role in the functions of self-powered tactile sensors and energy harvesters. Among these piezoelectric materials, ZnO has shown great potential due to its biocompatibility and easy synthesis in nano-structural forms and combined with properties like wide and direct bandgap 3.4 eV and large exciton binding energy of 60 meV, it can be utilized in many other applications.[12] ZnO comes with the advantages of easy synthesis and biocompatibility so it can play an important role in the development of self-powered and biocompatible nano/robotics haptics devices and energy harvester devices. Experimentally, the growth of ZnO nanostructures/nanorods (NRs) can be done with various simple and low-cost chemical-based synthesis approaches, but the control of the length, width, density, and verticality of ZnO NRs is a cumbersome task. There are various growth possibilities during synthesis in which ZnO NRs may be vertical, tilted, hollowed (nanotubes),[13] pointy (tip) type,[14] and also pyramid like-shape.[15] The various growths can also be seen on a single substrate, which can lead to a great variation in the piezoelectric potential due to the growth of various geometries, orientations, and sizes. So, it is more beneficial for the scientific community to quantitatively understand the effect of different growth geometry on its piezoelectric output. Therefore, in this paper, various geometries, sizes, and orientations are simulated with the finite element method (FEM) in COMSOL Multiphysics software to see its effect on piezoelectric potential. Many research groups have already studied and reported the effects of applied force on the vertically grown ZnO NRs, but the geometry and orientation-dependent piezoelectric response in ZnO NRs is still to be explored. [16, 17] The different geometries like the tilted, hollowed, and pyramid like-shape, ZnO NRs require attention which are rarely been reported. Therefore, in this paper, direct piezoelectric behavior in the different orientations and geometries of ZnO NRs have been computed by simulation.

## 2.   Basics of the direct piezoelectric effect

The piezoelectric potential can be understood in the terms of the geometrical stress confinement along various axes of a ZnO NR which results in the contraction or expansion of the dimension values along the axes. **Figure 1** shows the schematic diagram of the direct piezoelectric effect and pictorial representation of the direction of forces.

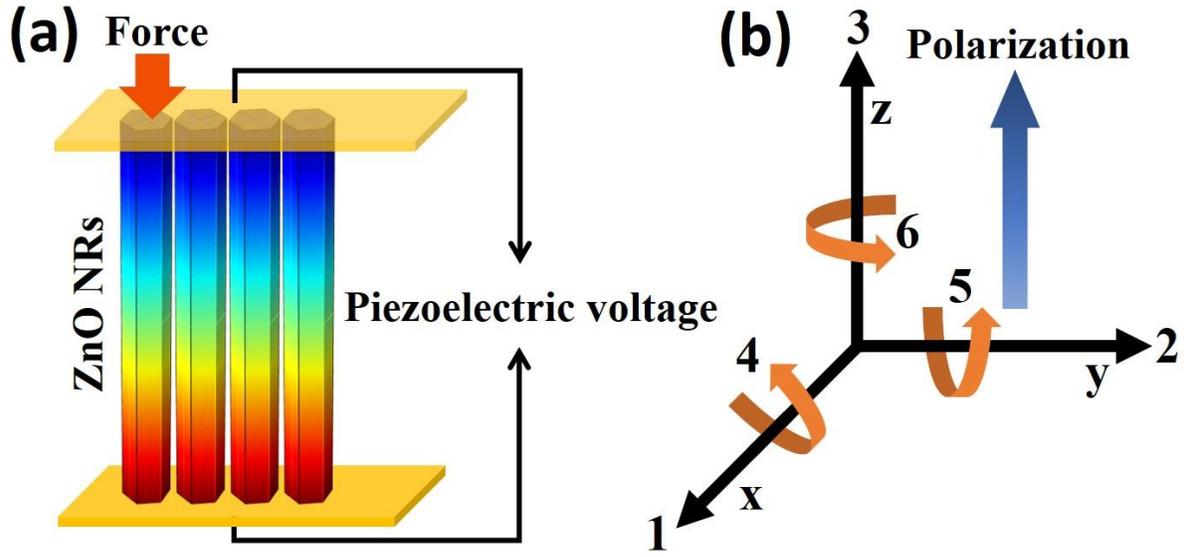

**Figure 1. (a)** Schematic illustration of the direct piezoelectric effect on a ZnO NRs based device and **(b)** Pictorial representation of the direction of forces on a piezoelectric material and polarization.

In the direct piezoelectric effect, there are several linear proportionality constants that connect the mechanical and electrical states and are defined as the piezoelectric coefficient (constant) (*d*). When stressed the piezoelectric material develops an electric charge (generally anisotropic), and an electrical potential is generated depending on the values of the piezoelectric coefficients as shown in Figure 1 (a). Hence, the equation that relates the mechanical and electrical variables can be written as (where *i, k* = 1,2,3 and *j* = 1, 2, ……6).[18]

$$D_i = d_{ij} T_j + \varepsilon_{ik} E_k \tag{1}$$

where $D_i$ is the electric displacement (C m$^{-2}$), $d_{ij}$ is the piezoelectric coefficient or piezoelectric strain coefficient for direct effect (C N$^{-1}$), $T_j$ is the mechanical stress (N m$^{-2}$), $\varepsilon_{ik}$ is the permittivity at constant stress (F m$^{-1}$), and $E_k$ is the electric field strength (V m$^{-1}$). Equation (1) shows the direct piezoelectric effect and the generation of electrical energy by the application of mechanical energy. The axes x, y, and z are indicated by the subscripts 1, 2, and 3, respectively, and the shear is indicated by subscripts 4, 5, and 6, respectively, the direction of polarization is taken in the z-direction, as can be seen in Figure 1 (b). The first subscript *i* represents in the piezoelectric coefficient notation $d_{ij}$, is the direction of induced polarization and the second subscript *j* represents the direction of applied stress.

## 3. Finite element approach for simulating piezoelectric ZnO NR

In case of complicated partial differential equations which cannot be solved analytically or lack exact solutions due to complex boundary conditions or functions, it is better to solve that mathematical problem with the help of FEM, which can provide an appropriate and realistic numerical solution. In FEM, the geometric system is divided as closely as possible into small and simpler domains, these domains are called the finite element, and these elements are connected at the points, and are called the nodes. The collection of the elements is said to be the finite element mesh, when the elements of the system are of the same dimension it is said to be uniform mesh, otherwise non-uniform mesh. To calculate the effect of stress on the piezoelectric potential of ZnO NR structures, FEM can be applied due involvement of multiple partial differential equations. The electromechanical constitutive equations for linear piezoelectricity are given by [19]

$$T = c^E S - d^t E \tag{2}$$

$$D = dS + \varepsilon^s E \tag{3}$$

Where $T$ is the mechanical stress vector (N m$^{-2}$), $c^E$ is the elasticity matrix or stiffness (N m$^{-2}$) (evaluated at the constant electric field), $E$ is the electric field vector (V m$^{-1}$ or N C$^{-1}$), $d$ is the piezoelectric stress coupling coefficient matrix (C m$^{-2}$) and the superscript $t$ denotes the transpose, $D$ is the vector of electric displacement (C m$^{-2}$), $S$ is the strain vector (unit less), and $\varepsilon^s$ is the dielectric matrix (F m$^{-1}$ or C V$^{-1}$ m$^{-1}$) (evaluated at constant mechanical strain). Equations (2) and (3) show the piezoelectric behavior of the material for which the FEM model is solved in the COMSOL software. Details of FEM to calculate the piezoelectric response with simulation can be found elsewhere.[20]

## 4. Types of ZnO NRs in the experimental synthesis

Various shapes and orientations are possible in the growth process of ZnO NRs. Many factors can be optimized to achieve vertical ZnO NRs but the main aim is to develop high piezoelectric potential generating devices. [21] In most of the ZnO NRs growth processes, conducted without much optimization have a tendency to develop tilted structures, there are also other possibilities due to changes in growth parameters viz. hollowed, pyramid-shaped, and with different aspect ratios. [22] **Figure 2** shows scanning electron microscope (SEM) images of various ZnO NR growths obtained with the seedless hydrothermal method. A low-cost seedless simple hydrothermal method was used for the growth of ZnO NRs on Au/Si substrate. Both Zinc nitrate hexahydrate and HMT (in a suitable amount with 1:1 ratio) were dissolved separately in DI water and mixed using a magnetic stirrer for 4-5 min to get a homogenous and well-mixed solution. The substrates were kept face down in a solution-filled glass beaker with a cap and placed in a preheated oven at 90$^0$C temperature for 16 hrs. After completion of growth, the substrates were rinsed thoroughly in hot (50-60$^0$C) DI water and dried with a nitrogen gun followed by baking on a hot plate for 3-4 min at 100$^0$C temperature.

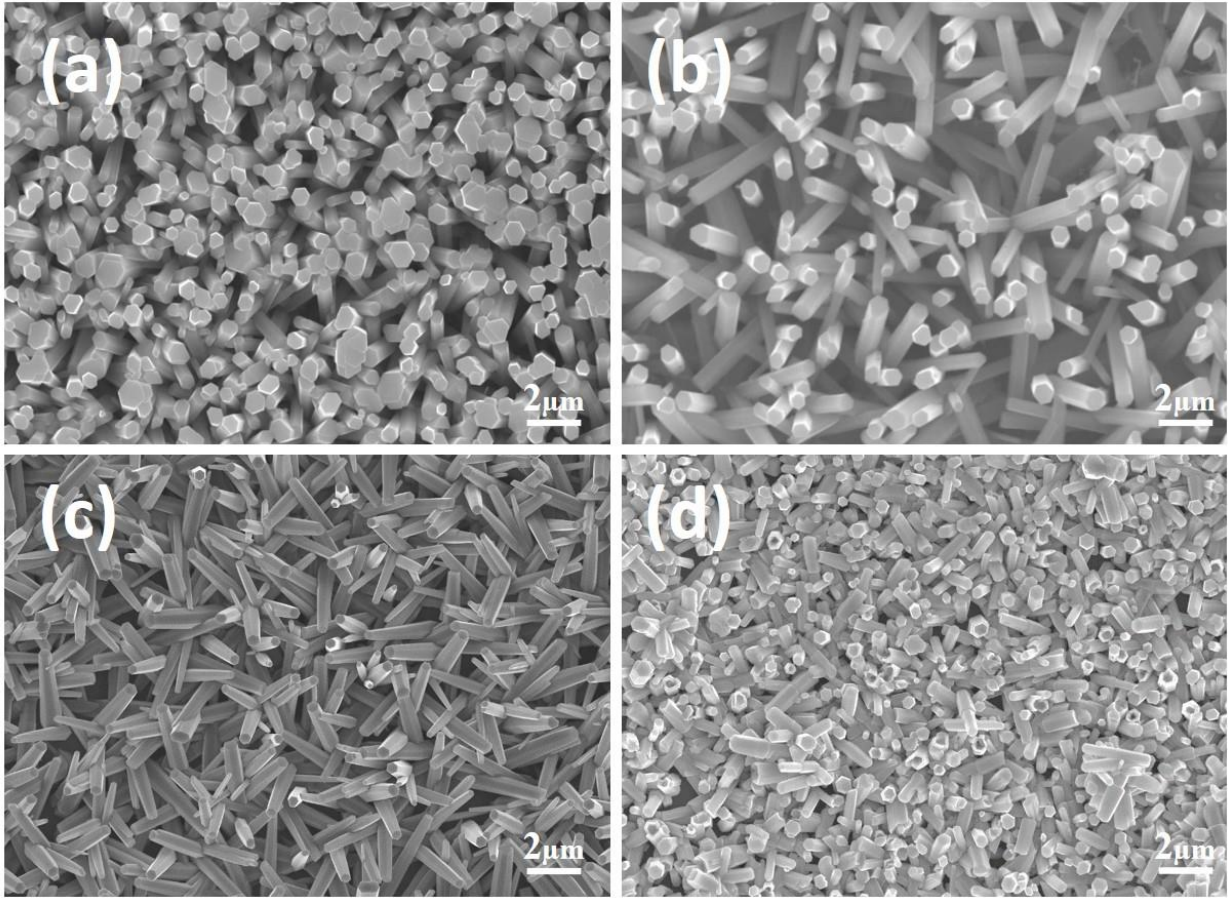

**Figure 2**. Various shapes and orientations of ZnO NRs obtained with the seedless hydrothermal growth method; **(a)** vertical, **(b)** tilted, **(c)** pyramid, and **(d)** hollow.

## 5. Result and discussions

### 5.1 Effect of aspect ratio on the piezoelectric potential of vertical ZnO NRs

The piezoelectric potential is first calculated for various dimensions of a vertical ZnO NR and later for various geometries and inclinations using simulation. The $c$-axis oriented ZnO NR is assumed to be intrinsic, and the bottom part of NR is fixed and electrically grounded. The uniformly applied force on the top surface of the NR is assumed to be loaded on the entire top hexagonal area of the NR, with other parameters like the piezoelectric coefficients and the mass density of ZnO NRs kept fixed.

The *c*-axis piezoelectric coefficient $d_{33}$ value is the highest in ZnO NRs so researchers in past were more interested in vertical growth to achieve the highest piezoelectric potential output. So, the *c*-axis oriented (002) ZnO NRs are preferred over any other growth for the development of high performance pressure/haptic sensors or energy harvesting devices. Experimentally it was found that the vertical orientation is possible with several optimized parameters but still there is a possibility of changes in the sizes (length and diameter). [23, 24] Hence, it is important to investigate the piezoelectric output voltage versus the size of the ZnO NRs by computational simulation. First, the structure is drawn with a 3D model of a *c*-oriented vertical hexagonal shaped ZnO NR in the simulation software with length of 3 µm and diameter of 500 nm. A pressure (boundary load) of 50 kPa is applied on the top of the NR. In electrostatics boundary conditions, the bottom part of the NR is electrically grounded and a floating potential is calculated on the top of the NR, and a value of 15.52 mV piezoelectric voltage is recorded in the simulation result as shown in **Figure 3 (a)**, which shows the 3D modeling and output voltage developed on a ZnO NR. Figure 3 (b) shows the von Mises stress and the strain on the ZnO NR.

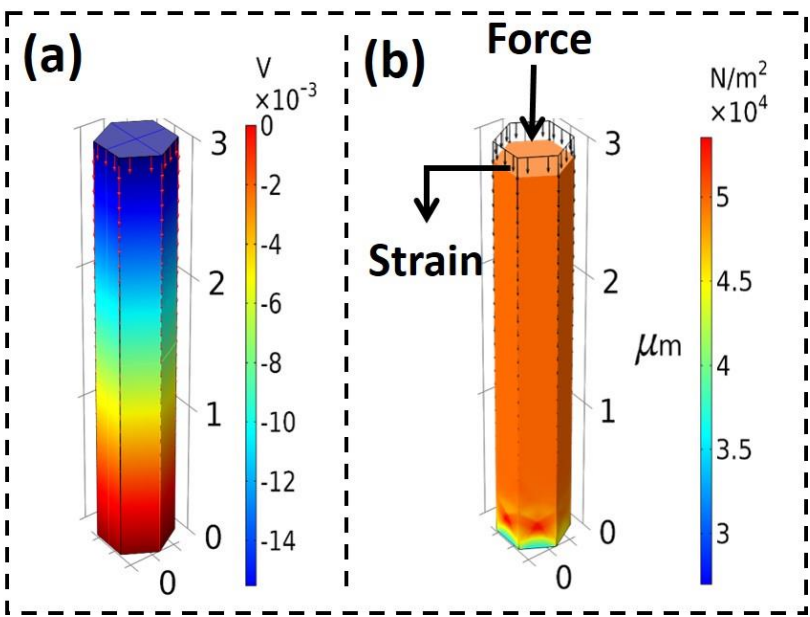

**Figure 3.** Simulated 3D model of single ZnO NR with (a) generated output voltage, and (b) von Mises stress and strain, in different sections at 50 kPa boundary load on top.

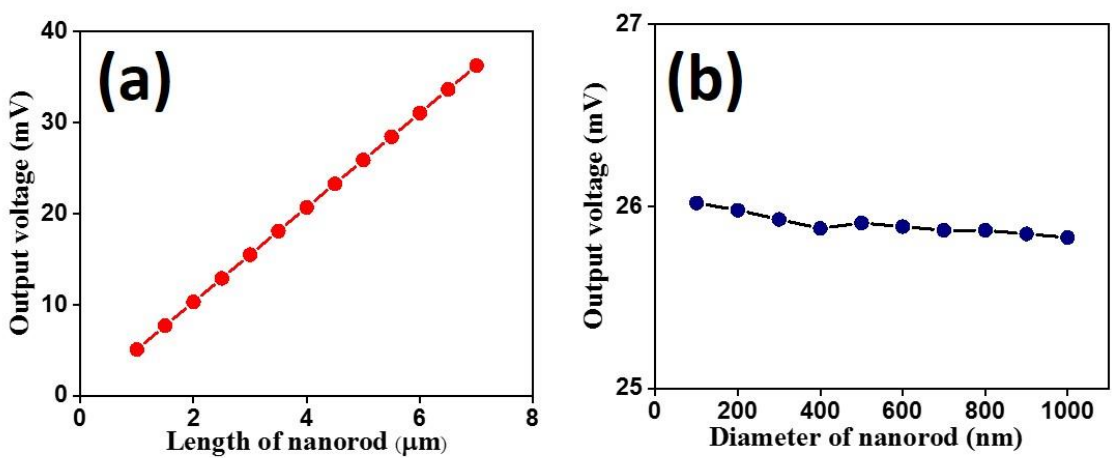

**Figure 4.** Simulated piezoelectric output voltage versus **(a)** the length (at a fixed diameter of 500 nm) and **(b)** diameter (at a fixed length of 5 µm) under a fixed applied stress of 50 kPa on a single ZnO NR.

The simulated output voltage of 15.52 mV has also been verified by numerical calculation value which is 15.64 mV for same pressure and dimensions of the nanorod. Furthermore, to understand the effect of size on the piezo potential, different aspect ratios (length/diameter) have been simulated. First, different lengths of ZnO NRs from 1 to 7 μm with a fixed diameter of 500 nm are simulated on the application of 50 kPa pressure on the top area of the NR. **Figure 4 (a)** shows the variation of output voltage generated vs the lengths of NR, as expected the voltage increases on increasing the length. In the second simulation, the effect of variation in diameter from 100 to 1000 nm on output voltage is simulated with a fixed length of 5 μm and under the same pressure of 50 kPa, and the result is shown in Figure 4 (b). The simulation results suggest that the piezoelectric voltage increases on increasing the length and decreases on increasing the diameter of the NR. The results from Figure 4 (a) and (b) are essentially related to the aspect ratio of the ZnO NR and its piezoelectric output, both results show that a higher aspect ratio leads to higher piezoelectric potential due to higher strain developed in higher lengths. An increase in the diameter leads to stress distribution across higher area leading to less strain and hence less piezoelectric potential development.

The input pressure can also linearly enhance the piezoelectric potential due to the much higher strain produced at higher stress values. Simulations based on stress values vs piezoelectric output are also carried out with keeping the length and diameter constant at 5 μm and 500 nm, respectively. **Figure 5** shows the effect of the pressure on the piezoelectric voltage which matches with the reported results. [25, 26]

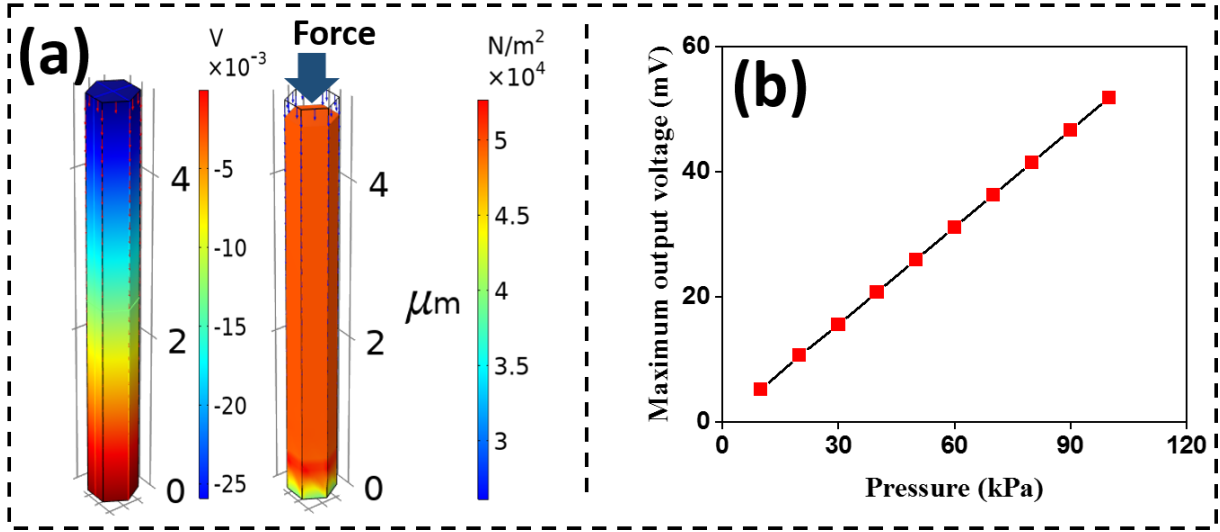

**Figure 5.** (a) Simulated 3D model of a ZnO NR with length and diameter, 5 μm and 500 nm respectively at a pressure of 50kPa, and (b) Generated piezoelectric output voltage versus the applied pressure.

## 5.2 Effect of pyramid-like NR on piezoelectric potential

Often chemical growth process of ZnO NRs leads to the growth of pyramid-like hexagonal structures in which the lower end of the NR is much wider relative to the top end.[27,28] In this research work the main aim is to find out the orientation and morphology of ZnO NR for maximum piezoelectric output, so simulations on pyramid-like hexagonal structures are carried out. A ZnO NR of length 5 μm with a fixed bottom cross-section area ($A_B$ = 0.2165 μm$^2$) or a fixed diameter (500 nm) is modeled in this simulation, to analyze the piezoelectric output voltage changes with varying top area ($A_T$), kept at a constant pressure of 50 kPa on the top.

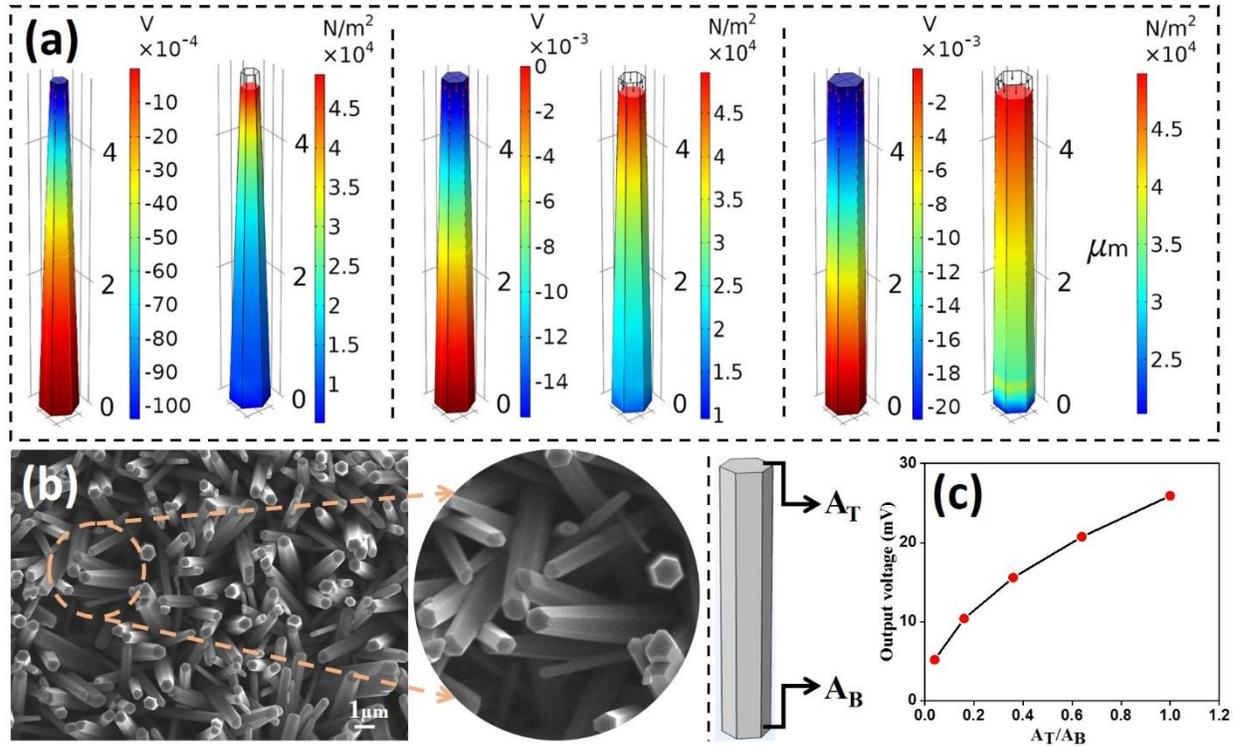

**Figure 6.** **(a)** Various simulated geometries (voltage and von Mises stress) of ZnO NRs with varying top cross-section area ($A_T$) and the fixed bottom area ($A_B$), **(b)** SEM image of hydrothermally grown ZnO NR with pyramid-like structures and a zoom view, and **(c)** variation of piezoelectric output voltage vs the area ratio ($A_T/A_B$) under the constant pressure of 50 kPa and fixed length 5 µm.

**Figure 6 (a)** shows the simulated response of the generated piezoelectric output voltages and von Mises stress with strain under a constant pressure of 50 kPa on top in three different values of $A_T$. The voltage scales demonstrate a rise in the output voltage with the increase in the top area, which could be due to higher volume available for generating voltage in case of higher $A_T$, the result also agrees with the reported findings.[27] Figure 6 (b) shows scanning electron microscope (SEM) images of pyramid type ZnO NRs growth with the hydrothermal growth process and Figure 6 (c) shows the effect of five different $A_T/A_B$ on the generated piezoelectric potential which demonstrate a rise in the output voltage with increase in $A_T$ values. The simulation results show that the microneedle (narrow pyramid-like shape) morphology of ZnO NR doesn't favor the production of the maximum output voltage, and hence not suitable for the development of high performance nano-generator or haptic sensors.

### 5.3 Effect of hollow (nanotubes) NR on piezoelectric potential

Hollow ZnO NRs growth is also very common in hydrothermal growth and hence must be investigated for the piezoelectric output generation.[29] In this section simulations are carried out to investigate the hollow ZnO nanotubes, a hollow NR of length 5 µm and outer diameter 500 nm is modeled with a varying inner diameter from 300 to 400 nm at a fixed pressure of 50 kPa.

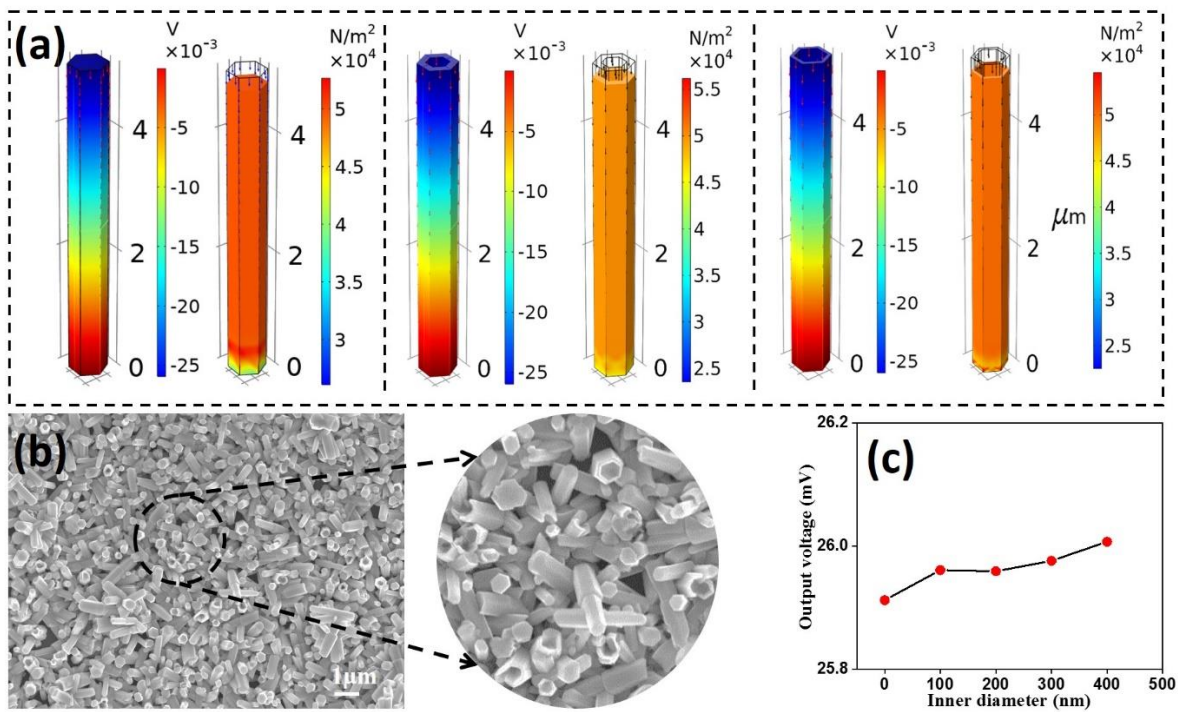

**Figure 7. (a)** Simulation results (piezoelectric potential, von Mises stress and strain) of solid and two hollow ZnO NRs with inner diameters of 300 and 400 nm and at a fixed outer diameter of 500

nm, **(b)** SEM images of hollowed ZnO NRs formed with the hydrothermal growth process, and **(c)** the magnitude of generated piezoelectric voltage with different values of inner diameter.

**Figure 7 (a)** shows the model of a solid and two hollow ZnO NR with two different inner diameters, with induced piezoelectric potential and von Mises stress are calculated along the vertical $c$-axis. The simulations show a minor change in the generated piezoelectric potential. Figure 7 (b) shows the SEM image of hollow ZnO NRs formed in the hydrothermal growth process and is also reported in the literature.[29] Figure 7 (c) shows the piezoelectric output variation with the inner diameter starting from solid to 400 nm, which displays minor enhancement in the output voltage, and also agrees with the previously reported results.[18] This minor enhancement in the piezoelectric potential for higher stress values is because of the decrease in the volume responsible for generating the output, so as higher strain is achieved at higher stress values at the same time the volume responsible for generating the output is also less, so overall it is a very small change in the generated output voltage.

## 5.4 Effect of tilt on piezoelectric potential

There have been many efforts in the growth of vertical ZnO NRs due to their higher value of piezoelectric coefficient along the $c$-axis. The growth of vertical and well-aligned ZnO NRs is possible with many optimized growth parameters in physical or chemical methods, a slightly un-optimized growth parameters generally have a tendency for tilted growth. Simulations are carried out to understand the effect of various tilt angles of the ZnO NRs on piezoelectric potential. Starting from a vertical single ZnO NR with a fixed length and diameter of 5 µm and 500 nm respectively is modeled to calculate the output voltage, and later the NR is given tilt from 30-90º angles from the surface of the substrate. The tilted geometry of ZnO NRs causes unsymmetric strain/stress distribution due to the shear stress on the top surfaces of NR. In the case of vertical ZnO NR, the pressure is applied only on the top surface area, but in the case of the tilted NRs, the pressure is experienced by the entire body exposed to the top side i.e. force is experienced by the top flat area and side facets of the NR which are exposed to the force. **Figure 8 (a)** shows simulation model images of ZnO NR with three different inclination angles viz. 40, 60 and 80° w.r.t. the substrate and both generated output voltage and von Mises stress with strain. Simulation is carried out for an applied pressure of 50 kPa on all the top facing areas of the NR. Figure 8 (b) shows the SEM image of hydrothermally obtained ZnO NRs, the image shows vertical as well as tilted NRs which are naturally obtained in any growth process of ZnO NRs. Figure 8 (c) shows the simulated tilted ZnO NR image and the inclination angle chosen from the surface of the substrate. Figure 8 (d) shows the variation of generated piezoelectric potential with different angles of tilt. The magnitude of the piezoelectric potential initially increases from angle 30 to 60° and then starts decreasing till the vertical alignment of NR. The higher values of generated piezoelectric potential indicate much higher stress levels in tilted ZnO NRs with a peak value of 216.21 mV at a 60° angle. The results are contrary to the expected results possibly due much higher level of stress experienced in tilted NRs due to shear stress and contributions from various piezoelectric coefficients. The results can be understood in terms of optimized contributions of both shear stress and piezoelectric coefficient along c-axis. The highest value of piezoelectric coefficient is along the c-axis but at 90° or vertical alignment the value of stress levels are much less due to the normal or compressional stress, which results in lower strain and hence lower values of generated output voltage. The shear stress increase with tilt and hence the value of generated output voltage increases with it and at the same time the contribution from piezoelectric coefficient along c-axis also reduces. The optimized value is achieved at 60° tilt where both shear stress and c-axis piezoelectric coefficient contributes to the maximum generated output voltage. The value of generated voltage decreases below 60° tilt angle due to much less contribution from c-axis piezoelectric coefficient despite the rise in shear stress values. The results show the devices made from tilted ZnO NR offer higher values of generated piezoelectric potential and hence this type of growth can be more useful for the development of high performance energy harvesting and haptic devices.

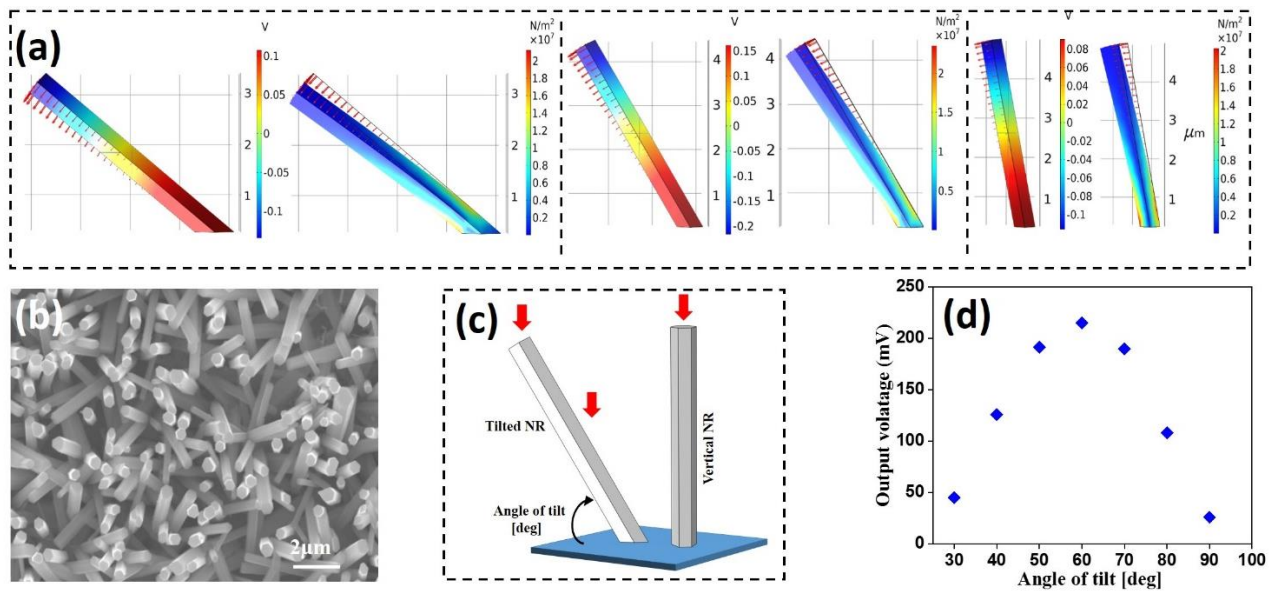

**Figure 8**. (**a**) Simulation results (piezoelectric potential, von Mises stress and strain) of ZnO NR with three different tilt angles of 40, 60 and 80° and at fixed diameter, length and pressure of 500 nm, 5µm, and 50kPa, respectively, (**b**) Experimental obtained tiled growth of ZnO NRs with the hydrothermal process, (**c**) Schematic representation of tilted geometry and pressure direction, and (**d**) The variation of piezoelectric output voltage with the tilt angle of ZnO NR at a constant pressure of 50 kPa.

## Conclusion

In conclusion, we have investigated the correlation between the size, shape, geometry, and tilt on the generated piezoelectric potential of ZnO NRs using COMSOL simulation to find the best morphology and shape for the development of high performance energy harvesters and haptic devices. The morphology and shape are chosen from the experimentally obtained ZnO NRs from the hydrothermal growth process. The simulation results show that the vertical NRs having a higher aspect ratio exhibit enhancement in the generated output potential for the same applied force. The morphology with the less top area also known as pyramid-like shape shows a decrease in generated piezoelectric potential and the hollow ZnO NRs show a minor increment with the hollowness of the NR. The highest generated piezoelectric potential is achieved in the case of tilted ZnO NRs with a maximum generated piezoelectric potential of 216.21 mV at a $60^{\circ}$ angle from the substrate. The higher values of piezoelectric potential are attributed to the optimized contributions from both higher shear stress level and c-axis piezoelectric coefficient contribution. Comparing the results with the actual obtained morphology during the hydrothermal growth process where the tendency of tilted growth is much higher, suggest that the naturally obtained tilted growth morphology are best suited for the development of piezoelectric energy harvesters and haptic device. This result is contrary to the expected results, according to which the vertically aligned ZnO NRs should produce the highest piezoelectric output voltage.

## Author Contributions

Rehan Ahmed: data curation, investigation, formal analysis, visualization, writing original draft, writing-review, and editing. Pramod Kumar: conceptualization, methodology, formal analysis, visualization, funding acquisition, project administration, resources, supervision, writing-original draft, writing-review, and editing.

**Conflicts of interest**

There are no conflicts to declare.

**Acknowledgments**

The authors would like to thank the Indian Institute of Technology Bombay (IITB) for the seed grant and the Science and Engineering Research Board (SERB) Government of India for funding. The authors would like also to thank the Council of Scientific and Industrial Research (CSIR) Government of India for providing the fellowship.